\documentclass[aps, prb, floatfix, twocolumn, superscriptaddress]{revtex4-2}

\usepackage{xcolor}
\usepackage{mathrsfs}
\usepackage{graphicx}
\usepackage[english]{babel}
\usepackage{amsmath}
\usepackage{amssymb}
\usepackage[normalem]{ulem}

\usepackage{relsize}
\usepackage{CJK}
\usepackage{dsfont}
\usepackage{bm}% bold math

\newcommand{\me}{\mathrm{e}}
\newcommand{\mi}{\mathrm{i}}

\newcommand{\dif}{\mathrm{d}}

\allowdisplaybreaks

\begin{document}

\title{Mixed-state geometric phases of coherent and squeezed spin states}

\author{Xin Wang}
\affiliation{School of Physics, Southeast University, Jiulonghu Campus, Nanjing 211189, China}

\author{Jia-Chen Tang}
\affiliation{School of Physics, Southeast University, Jiulonghu Campus, Nanjing 211189, China}

\author{Xu-Yang Hou}
\affiliation{School of Physics, Southeast University, Jiulonghu Campus, Nanjing 211189, China}

\author{Hao Guo}
\email{guohao.ph@seu.edu.cn}
\affiliation{School of Physics, Southeast University, Jiulonghu Campus, Nanjing 211189, China}
\affiliation{Hefei National Laboratory, Hefei 230088, China}

\author{Chih-Chun Chien}
\email{cchien5@ucmerced.edu}
\affiliation{Department of Physics, University of California, Merced, California 95343, USA}

\begin{abstract}
Two mixed-state geometric phases, known as the Uhlmann phase and interferometric geometric phase (IGP), of spin coherent states (CSSs) and spin squeezed states (SSSs) are analyzed. 
Exact solutions and numerical results of selected examples are presented. For the $j = 3/2$ CSS, the Uhlmann phase exhibits finite-temperature topological phase transitions characterized by abrupt jumps. The IGP for the same state similarly shows discontinuous jumps as the temperature varies.
In the case of the $j = 1$ one-axis SSS, both Uhlmann phase and IGP display discrete finite-temperature jumps.
By contrast, the $j = 1$ two-axis SSS shows no such transitions because the Uhlmann phase and IGP both vary smoothly with temperature.
We also briefly discuss potential realizations and simulations related to these phenomena in spin systems.
\end{abstract}

\maketitle
\section{Introduction}
Geometric phases have been a key concept in topological properties of quantum systems~\cite{Bohm_GPbook,ChruscinskiBook}. The Berry phase arises when a quantum system undergoes adiabatic cyclic evolution in the context of pure quantum states and has been fundamental in studying topological insulators and superconductors, where the Berry curvature characterizes the underlying topological features~\cite{TKNN,Haldane,KaneRMP,ZhangSCRMP,KaneMele,KaneMele2,ChiuRMP,Bernevigbook,BernevigPRL,MoorePRB,FuLPRL,Bohm03,vanderbilt2018berry,cohen2019geometric}. However, quantum systems are often in mixed states due to finite temperatures or non-equilibrium conditions, which call for generalizations of the Berry phase to mixed quantum states. A mathematically rigorous approach is the Uhlmann phase~\cite{Uhlmann86,Uhlmann89,Uhlmann1992,Uhlmann91} for mixed states. Similar to the Berry phase associated with the holonomy of the Berry bundle, the Uhlmann phase is obtained from parallel transport in a cyclic process and reflects the holonomy of the Uhlmann bundle. Moreover, the Uhlmann-Berry correspondence shows that the Uhlmann phase in the low-temperature limit generally agrees with the Berry phase of the ground state~\cite{P2}. However, the Uhlmann bundle is trivial~\cite{TDMPRB15}, which limits its applicability to defining other quantized topological indices. Nevertheless, the Uhlmann phase of bosonic and fermionic coherent states have been studied~\cite{P2}, in addition to that of two-level and spin-$j$ systems~\cite{Viyuela14,P3,PhysRevA.103.042221}. With a suitable generalization, the Uhlmann phase of dynamic systems has also been studied~\cite{OurUhlmannQuench}.

Another approach called the interferometric geometric phase (IGP) for mixed states has been proposed~\cite{PhysRevLett.85.2845} and further analyzed~\cite{PhysRevA.67.020101,PhysRevA.70.052109,Faria_2003,Chaturvedi2004,PhysRevLett.93.080405,Kwek_2006}, extending the geometric phase concept via an analogy of the optical Mach-Zehnder interferometer~\cite{PhysRevLett.94.050401}. This IGP has been experimentally observed using nuclear magnetic resonance and polarized neutrons~\cite{PhysRevLett.91.100403,Ghosh_2006,PhysRevLett.101.150404,PhysRevLett.94.050401}. Unlike the Uhlmann phase, the IGP does not require the evolution of the system to be cyclic~\cite{PhysRevLett.85.2845}. In some cases, it can be shown that the IGP is the argument of the weighted sum of the Berry-phase factors of the constituent states~\cite{PhysRevLett.85.2845}.

Ref.~\cite{Hou2023} presents a comparison between the Uhlmann phase and IGP to contrast their differences. The Uhlmann phase, for instance, requires the ancillary system modeling environmental effects to be manipulated according to the change of the system and can exhibit finite-temperature topological phase transitions with quantized jumps in two-level systems. In contrast, the IGP does not exhibit such transitions at finite temperatures for general two-level systems. Recent studies have demonstrated that the IGP remains intact as temperature varies for the Kitaev chain~\cite{Andersson_2016}, in contrast to the discrete jumps seen in the Uhlmann phase \cite{Viyuela14}. Nevertheless, quantized jumps of the IGP as temperature increases can occur in three-level or higher-dimensional systems~\cite{Hou2023}.
Interestingly, these generalizations of the mixed-state geometric phases provide further insights into topological quantum phase transitions (TQPTs) and dynamical quantum phase transitions (DQPTs) \cite{OurPRB20b}.

Meanwhile, spin systems have served as paradigms in various physics branches, ranging from quantum mechanics~\cite{MQM} to statistical physics~\cite{HuangSM} to spintronic devices~\cite{CahaySpintronics}. Two types of spin states have received particular attention: The coherent spin states (CSSs) are minimal uncertainty states of collective angular momentum. The CSS is the building block for the path-integral description of many-body spin systems~\cite{Auerbach_GPbook} and serves as a classical approximation of quantum spin systems, thereby establishing the standard quantum limit (SQL) for measurements~\cite{PhysRevD.23.1693,Giovannetti_2004,PhysRevA.46.R6797,pezze2014quantum}. However, surpassing the SQL is critical for advancing quantum metrology and quantum information technologies. Spin squeezing~\cite{PhysRevA.47.5138,MA201189,Block2024} allows for a reduction of the variance of one spin component at the expense of increasing that of another orthogonal direction, resulting in the squeezed spin states (SSSs). The SSSs enable precision measurements beyond the SQL and approach the Heisenberg limit with potential applications in, for example, atomic clocks~\cite{PhysRevA.50.67,Eckner2023}. Significant improvements in precision measurement using spin squeezing have also been made in atomic Bose-Einstein condensates (BECs)~\cite{Est_ve_2008,Gross_2010}. These developments not only highlight the metrological advantage of the SSS but also open possible avenues for generating quantum entanglement in many-body systems, fostering further advancement in quantum information processing~\cite{Riedel_2010}.

Complementing the progress in surpassing the SQL through spin squeezing, here we set to investigate the geometric phases of the CSS and SSS at finite temperatures and search for topological phases or transitions. The possibility of finding different mixed-state geometric phases according to different evolution conditions already demonstrates rich physics in the underlying topology of finite-temperature systems. While the geometric phase of the ground-state CSS has been analyzed~\cite{Chaturvedi87,Chryssomalakos_2018}, here we explicitly calculate the Uhlmann phase and the IGP for selected CSS and SSS at finite temperatures. We found that for the $j=3/2$ CSS, both Uhlmann phase and IGP demonstrate temperature-dependent topological or geometric phase transitions with quantized jumps. For the $j = 1$ one-axis SSS, both phases show similar finite-temperature transitions. In contrast, the $j = 1$ two-axis SSS displays a smoothly varying Uhlmann phase with increasing temperature, indicating the absence of topological phase transitions. Likewise, the IGP of that case also evolves continuously with temperature.

The rest of the paper is organized as follows. Sec.~\ref{Sec.2} provides a brief, self-contained theoretical background, covering the CSS, SSS, Uhlmann phase, and the adiabatic IGP. 
Sec.~\ref{Sec.3} and Sec.~\ref{Sec.4} present our systematic investigations of the mixed-state geometric phases for selected cases of the CSS and SSS, respectively. Section \ref{Sec.5} discusses possible
experimental and theoretical implications of our findings. Finally, Sec.~\ref{Sec.6} concludes our work.

\section{THEORETICAL BACKGROUND}\label{Sec.2}
We begin with a brief overview of the CSS and SSS as well as a short review of the Uhlmann phase and interferometric geometric phase (IGP) of mixed states. Additionally, we will generalize the concept of IGP to allow arbitrary unitary evolution.
For simplicity, we set $c = \hbar = k_\text{B} = 1$ throughout the paper.

\subsection{Overview of CSS}
The spin operators satisfy the algebra \cite{MQM}
\begin{align}\label{liebracket}
	[J_z,J_+]=J_+,\quad [J_z,J_-]=-J_-,\quad [J_+,J_-]=2J_z.
\end{align}
The eigenstates of $J^2$ and $J_z$ are $|jm\rangle,~-j\le m\le j$. Moreover, $J_\pm|jm\rangle=\sqrt{(j\mp m)(j\pm m+1)}|jm\pm1\rangle$ and $J_z|jm\rangle=m|jm\rangle$. The CSS is defined as \cite{Perelomov_1986,Radcliffe71}
 \begin{align}
	|\xi\rangle:=\me^{\xi J_+-\bar{\xi}J_-}|j,-j\rangle,
\end{align}
where $\xi\in \mathds{C}$ is a complex number.

Using the Baker-Campbell-Hausdorff disentangling formula~\cite{Perelomov_1986}, the CSS can also be expressed as
\begin{align}\label{De1}
	|\xi\rangle &=\mathrm{e}^{\zeta J_{+}} \mathrm{e}^{\ln \left(1+|\zeta|^{2}\right) J_{z}} \mathrm{e}^{-\bar{\zeta} J_{-}} |j,-j\rangle\notag\\
	&=\frac{1}{(1+|\zeta|^2)^j}\me^{\zeta J_+}|j,-j\rangle\equiv |\zeta\rangle
\end{align}
with $\zeta=\zeta(\xi)=\frac{\xi\tan|\xi|}{|\xi|}$.
After defining $D(\xi)\equiv D(\zeta)=\mathrm{e}^{\xi J_{+}-\bar{\xi} J_{-}}$, the CSS corresponds to the ground state of the translated Hamiltonian $\hat{H}(\xi)\equiv\hat{H}(\zeta)=D(\zeta)\hat{H}D^\dagger(\zeta)$.
Here $\hat{H}=\omega_0 J_z$ , and $\omega_0$ is the Larmor frequency.
The associated excited states are obtained in a similar manner:  $|jm,\xi\rangle=|jm,\zeta\rangle=D(\zeta)|jm\rangle$, $-j\le m\le j$.
By parameterizing $\xi$ as $\xi=\me^{-\mi\phi}\frac{\theta}{2}$, we have  $\zeta=\me^{-\mi\phi}\tan\frac{\theta}{2}$. If the parameter ranges are chosen as  $0\le \phi\le 2\pi$, $0\le \theta\le \pi$, then $\zeta$ spans the entire complex plane. Consequently, we will use $\zeta$ as the control parameter. The map $(\theta, \phi)\rightarrow \zeta$ projects the unit sphere onto the entire complex plane. Actually, the CSSs are closely related to the eigenstates of the corresponding spin-$j$ system with the Hamiltonian
$H=\omega_0 \hat{\mathbf{B}} \cdot \mathbf{J}$,
where $\hat{\mathbf{B}}$ is the unit vector pointing along the direction of an external magnetic field~\cite{Auerbach_GPbook}.

\subsection{Overview of SSS}
A spin-$j$ system can be regarded as a composite system consisting of 2$j$ spin-$\frac{1}{2}$ objects. The angular momentum operators for the collection of spin-$\frac{1}{2}$ objects are given by
\begin{align}
	J_\alpha=\frac{1}{2} \sum_{l=1}^N \sigma_{l \alpha}, \quad \alpha=x, y, z,
\end{align}
where $\sigma_{l \alpha}$ is the Pauli matrix for the $l$-th object. Here we consider the spin-squeezed states (SSS) generated by the one-axis twisting Hamiltonian \cite{MA201189,PhysRevB.76.064305}
\begin{align}\label{Eq:SSS}
	H_{\mathrm{OAT}}=\eta J_x^2=\frac{\eta}{4} \sum_{k, l=1}^N \sigma_{k x} \sigma_{l x},
\end{align}
which is a nonlinear operator with the coupling constant $\eta$ involving all pairwise interactions. This indicates the presence of pairwise correlations in the spin-squeezed states generated by the Hamiltonian. Choosing the initial state as $|j, -j\rangle=|1\rangle^{\otimes N}$, where $|1\rangle=|j=1/2,j_z=1/2\rangle$, the SSS at time $t$ is formally written as:
\begin{align}\label{SSS1}
	|\Psi(t)\rangle=\exp \left(-\mi \Theta(t) J_x^2 / 2\right)|1\rangle^{\otimes N},
\end{align}
where $\Theta(t)=2 \eta t$.

By defining
\begin{equation}\label{Eq:S}
S(\Theta)=\me^{-\mi \Theta J_x^2 / 2},
\end{equation} 
we can generalize this concept of spin-squeezed states 
%by acting the operator $\exp \left(-\mi \Theta J_x^2 / 2\right)$ on $|j m\rangle$:
\begin{align}
|jm,\Theta\rangle=S(\Theta)|jm\rangle.
\end{align}
We also introduce
\begin{align}\label{tildeS}
\tilde{S}(\Theta(t))=\sum_{m} \me^{-\int^{t}_{0} \dif t' \langle m(t') |\frac{\dif}{\dif t'} | m(t') \rangle}  | m(t) \rangle \langle m(0)|,
\end{align}
where $| m(t) \rangle:=|jm,\Theta\rangle$, so that $| m(0) \rangle=|jm\rangle$. 
When compared to $S(\Theta)$,  $\tilde{S}(\Theta)$ simply incorporates a pure-state geometric phase $\mi\mathlarger{\int}^{t}_{0} \dif t' \langle m(t') |\frac{\dif}{\dif t'} | m(t') \rangle$ to each eigenstate. It is straightforward to verify that 
$|jm,\Theta\rangle$ is an eigenstate of the Hamiltonian
\begin{eqnarray}\label{Eq:HTheta}
H(\Theta)&=& S(\Theta) \omega_0J_z S^\dag(\Theta) \nonumber \\
&=&\tilde{S}(\Theta(t)) \omega_0J_z \tilde{S}^\dag(\Theta(t))
\end{eqnarray}
since the geometric phase factors cancel out for each $|jm\rangle$. This in turn implies that the time-evolved thermal state satisfies $\rho(t)=S(\Theta(t))\rho(0)S^\dag(\Theta(t))=\tilde{S}(\Theta(t))\rho(0) \tilde{S}^\dag(\Theta(t))$. The advantage of introducing  $\tilde{S}$ will become clear in the subsequent discussions.

Another spin squeezing method is based on a generalized two-axis counter-twisting Hamiltonian \cite{PhysRevA.47.5138,MA201189}, which reads
\begin{align}\label{Eq:tSSS}
	H_{\mathrm{TAT}}=\mi (z J^2_+ -\bar{z} J^2_-).
\end{align}
Here $z\in \mathds{C}$ is a complex number which can be parameterized as $z=\me^{-\mi \phi } \tan \left(\frac{\theta }{2}\right)$ with the parameter ranges chosen as $0\le \phi\le 2\pi$, $0\le \theta\le \pi$. This is a quadratic squeezing Hamiltonian and it will degrade to the ordinary two-axis Hamiltonian $\frac{\gamma}{2\mi}(J_{+}^{2}-J_{-}^{2})$ if $z$ is a purely real number equal to $-\gamma/2$. The squeezing operator $K(z)$ now is defined as
\begin{equation}\label{Eq:K}
	K(z)=\me^{z J^2_+ -\bar{z} J^2_-},
\end{equation}
which is somewhat similar with the translation operator $D(\xi)=\mathrm{e}^{\xi J_{+}-\bar{\xi} J_{-}}$ we encountered in the discussion of the CSS. However, there are subtle differences in the quadratic form of angular momentum operators. Similar to the case of one-axis squeezing, the eigenstates $|jm,z\rangle$ can be given from the corresponding Hamiltonian
\begin{eqnarray}\label{Eq:Hztwoaixs}
	H(z)= K(z) \omega_0J_z K^\dagger(z).
\end{eqnarray}

\subsection{Uhlmann phase}
A mathematically rigorous approach for generalizing the Berry phase from pure states to mixed states is the Uhlmann phase \cite{Uhlmann86} through purification of density matrices, which can be described by the underlying fiber-bundle structure \cite{P1}, An operator $W$ is said to purify the density matrix $\rho$ if $\rho=WW^\dag$. Conversely, the purification $W$ can be uniquely decomposed as $W=\sqrt{\rho}\mathcal{U}$ if $\rho$ is full-rank (i.e., rank$(\rho)=N$ for a system with a $N$-dimensional Hilbert space). Here the unitary operator $\mathcal{U}$ is referred to as the phase factor of mixed states, representing a generalization of the U(1) phase factor associated with pure states.

We consider a system depending on a set of parameters $\mathbf{R}=(R_1,R_2,\cdots, R_k)^T\in M$, where $M$ is the parameter manifold. When $\rho$ evolves along a smooth curve $C(t):=\mathbf{R}(t)$ in $M$, the purification of the density matrix also continuously evolves along an induced curve $\gamma(t):=W(t)= \sqrt{\rho(\mathbf{R}(t))}\mathcal{U}(\mathbf{R}(t))$. If the length of $\gamma$, defined as $L(\gamma)=\mathlarger{\int}_\gamma\sqrt{\text{Tr}(\dot{W}\dot{W}^\dag)}\dif t$, is minimized, $W$ is said to undergo a parallel-transport process. Therefore, $\gamma(t)$ is the horizontal lift of $\rho(t)\equiv \rho(\mathbf{R}(t))$. It can be shown that the Uhlmann parallel-transport condition is given by \cite{Uhlmann95}
\begin{align}\label{pxcmU}
\dot{W}W^\dag=W\dot{W}^\dag.
\end{align}
This condition maximally ensures the ``parallelity'' between purification of adjacent states.

When $\rho$ experiences a cyclic evolution over a duration $\tau$, such that $\rho(0)=\rho(\tau)$, the evolution of its purification may not necessarily return to its initial value. Specifically, the initial and final phase factors of a Uhlmann process are related by
\begin{align}
\mathcal{U}(\tau)=\mathcal{P}\me^{-\oint_C A_U}\mathcal{U}(0),
\end{align}
where $\mathcal{P}$ is the path-ordering operator and
\begin{align}\label{AUE}
A_U=-\sum_{mn}|m\rangle\frac{\langle m|[\mathrm{d}\sqrt{\rho},\sqrt{\rho}]|n\rangle}{\lambda_m+\lambda_n}\langle n|
\end{align}
is the Uhlmann connection. Here $\lambda_n$, $|n\rangle$ represent the eigenvalues and eigenstates of $\rho$, respectively.
Notably, the parallelity lacks transitivity \cite{OurPRB20b} even during parallel transport. Hence, the operator $\mathcal{P}\me^{-\oint_C A_U}$ associated with the holonomy of the Uhlmann bundle quantifies a measure of the loss of parallelity between the initial and final purifications. The following Uhlmann phase then reflects the topological change of the Uhlmann holonomy.
\begin{align}\label{thetaU}
\theta_U(C)=\arg\text{Tr}(W^\dag(0)W(\tau))=\arg\text{Tr}[\rho(0)\mathcal{P}\me^{-\oint_C A_U}].
\end{align}

\subsection{Interferometric geometric phase}
There is another approach for constructing a geometric phase for mixed states. Inspired by optical processes in the Mach-Zehnder interferometer, Ref.~\cite{PhysRevLett.85.2845} assigned a phase to a mixed quantum state represented by the density matrix $\rho$ undergoing a unitary evolution $U(t)$.
Explicitly, if a system evolves according to $\rho(t)=U(t)\rho(0)U^\dagger(t)$, it gives rise to a total phase
\begin{align}\label{thetaT}
\theta_\text{T}(t)=\arg\text{Tr}\left[\rho(0)U(t)\right].
 \end{align}
Notably, the transformation from $\rho(t)$ to $\rho(t+\dif t)$ is $\rho(t+\dif t) =U(t+\dif t)\rho(0)U^\dagger(t+\dif t)=U(t+\dif t)U^\dagger (t)\rho(t)U(t)U^\dagger(t+\dif t)$. Thus, by definition, the relative phase between the mixed states at $t$ and $t+\dif t$ is $\theta_\text{T}(t)=\arg\text{Tr}\left[\rho(t)U(t+\dif t)U^\dagger (t)\right]$.
In this framework~\cite{PhysRevLett.85.2845}, a parallel-transport condition of $U(t)$ is proposed as follows.  $\rho(t+\dif t)$ is required to be instantaneously ``in phase'' with $\rho(t)$, which means  $\arg\text{Tr}\left[\rho(t)U(t+\dif t)U^\dagger (t)\right]=0$. In the differential form, the condition yields
\begin{align}\label{pxcm1}
	\text{Tr}\left[\rho(t)\dot{U}(t)U^\dagger (t)\right]=\text{Tr}\left[\rho(0)U^\dag(t)\dot{U}(t)\right]=0.
\end{align}
To specify the parallel transport by the evolution according to $U(t)$ more deterministically, the condition (\ref{pxcm1}) has been strengthened as \cite{PhysRevLett.85.2845}
\begin{align}\label{pxcm2}
\langle n(t)|\dot{U}(t)U^\dagger (t)|n(t)\rangle=0,\quad n=0,1,2,\cdots,
\end{align}
where $|n(t)\rangle$ is the $n$th eigen-vector of $\rho(t)$. We emphasize that the parallel-transport condition (\ref{pxcm1}) or its strengthened form \eqref{pxcm2} for the IGP fundamentally differs from Uhlmann's condition in Eq. (\ref{pxcmU}), due to the U$(N)$ gauge dependence of the phase factor $\mathcal{U}$. Crucially, introducing $\mathcal{U}$ via $W = \sqrt{\rho} \mathcal{U}$ can violate the parallel-transport condition (\ref{pxcm1}) \cite{OurTB}.

In general, the IGP parallel-transport condition shown in Eq.~\eqref{pxcm2} can always be satisfied, as demonstrated in Refs.~\cite{PhysRevLett.85.2845,PhysRevLett.93.080405}.
For instance, in our earlier discussion of the one-axis SSS, we introduced the operator
$\tilde{S}$. It can be shown that for all $n$,
\begin{align}\label{tildeSptc}
&\langle n(t)| \dot{\tilde{S}}(t)\tilde{S}^\dagger(t) | n(t) \rangle = \langle n(t)| \sum_{m} \frac{\dif |m(t)\rangle}{\dif t} \langle m(t)| n(t) \rangle \notag\\
-& \langle n(t)| \sum_{m} \langle m(t)| \frac{\dif}{\dif t} |m(t)\rangle |n(t)\rangle \delta_{mn} = 0.
\end{align}
Hence, the IGP parallel-transport condition is automatically satisfied, regardless of the explicit form of $S$.

Under the parallel-transport condition, the phase accumulated during the evolution is the geometric phase referred to as the IGP:
\begin{align}\label{GPm}
	\theta_\text{G}(t) =\arg\text{Tr}\left[\rho(0)U(t)\right].
\end{align}
This is because there is no accumulation of the dynamic phase during parallel transport, which can be understood by noting that if $U(t)$ represents a dynamic evolution process according to $\mi\dot{U}=HU$, the parallel-transport condition (\ref{pxcm1}) guarantees
\begin{align}\label{Pdyn}
	\theta_\text{D}(t)&=-\int_0^t\dif t'\text{Tr}\left[\rho(t')H(t')\right]\notag\\&=-\mi\int_0^t\dif t'\text{Tr}\left[\rho(t')\dot{U}(t')U^\dag(t')\right]=0.
\end{align}
As a consequence, the total phase contains only the geometric contribution from the IGP. We note that the strengthened condition~\eqref{pxcm2} will result in vanishing integrand in the above expression, thereby resulting the same vanishing dynamic phase.
We mention that general unitary evolution beyond parallel transport could result in non-adiabatic phases~\cite{ZHANG20231}.
There is a formalism known as the quantum kinematic approach to the geometric phase for pure states~\cite{MUKUNDA1993205}, and a similar non-adiabatic geometric phase for mixed states has been discussed in Ref. \cite{GPMS24}.

\subsection{Difference between the two geometric phases}
We emphasize that the two aforementioned geometric phases for mixed states are physically and mathematically distinct. Although the theory of IGP can also be formulated equivalently using a purification formalism of density matrices \cite{PhysRevLett.85.2845}, any phase factor $\mathcal{U}$ is ruled out in the derivation. Specifically, only the trivial decomposition $W=\sqrt{\rho}$ is permissible for the IGP because otherwise the condition (\ref{pxcm1}) would be violated.

In contrast, the generation of the Uhlmann phase inherently relies on the phase factor $\mathcal{U}$ according to the underlying fiber-bundle structure. All purifications of a given density matrix $\rho$ thus span a U$(N)$ fiber space. Consequently, the Uhlmann phase is naturally and directly connected to the topological properties of the U$(N)$ principal bundle, a feature absent in the IGP. Explicitly, the Uhlmann phase corresponds to the Uhlmann holonomy of the Uhlmann bundle of density matrices, similar to the Berry phase revealing the Berry holonomy of the fiber bundle of pure states~\cite{P1}.
Moreover, the Uhlmann phase requires a cyclic process but is guaranteed to be gauge invariant~\cite{PhysRevLett.85.2845,Hou2023} while the IGP and its generalization do not impose those constraints.
In the following, we will compare the differences between the Uhlmann phase and IGP through specific examples of the CSS and SSS.

\section{Mixed-state Geometrical Phases of CSS}\label{Sec.3}

\subsection{The Uhlmann phase}
To evaluate the Uhlmann phase, we consider a loop in the complex $\zeta$ plain: $C(t)\equiv\zeta(t)$, $0\le t\le \tau$ with $\zeta(0)=\zeta(\tau)$. The continuous transformation $D(\zeta(t))$ of a CSS at temperature $T$ induces a corresponding loop  $\gamma(t):=\rho(\zeta(t))$ in the manifold of density matrices, where
\begin{align}\label{rho1}
	\rho(\zeta)=\frac{1}{Z}\me^{-\beta\hat{H}(\zeta)}=\frac{1}{Z}D(\zeta)\me^{-\beta\hat{H}}D^\dag(\zeta).
\end{align}
Here $\beta=1/(k_B T)$. Since $D(\zeta) $ is unitary, the eigenvalues of the Hamiltonian remain invariant under $D(\zeta)$. As a result, the partition function is a constant: $Z=\sum_{m=-j}^j\me^{-m\beta\omega_0}=\frac{\sinh\left[\left(j+\frac{1}{2}\right)\beta\omega_0\right]}{\sinh\left(\frac{\beta\omega_0}{2}\right)}$.

According to Eq.~(\ref{AUE}), the Uhlmann connection of the CSS is given by
\begin{align}\label{AusHo1}
	&A_U=-\sum_{n\neq m}\frac{(\sqrt{\lambda_n}-\sqrt{\lambda_m})^2}{\lambda_n+\lambda_n}| jn,\zeta\rangle\langle jn,\zeta|\dif|jm,\zeta\rangle\langle jm,\zeta |\notag\\
	&=-\sum_{n\neq m}\chi_{nm}D(\zeta)
	|jn\rangle\langle jn|D^\dag(\zeta)\dif D(\zeta)|jm\rangle\langle jm|D^\dag(\zeta),
\end{align}
where $\lambda_m=\frac{1}{Z}\me^{-m\beta\omega_0}$ is the $m$th eigenvalue of the density matrix, and $\chi_{nm}=\frac{(\me^{-\frac{n}{2}\beta\omega_0 }-\me^{-\frac{m}{2}\beta\omega_0 })^2}{\me^{-n\beta\omega_0 }+\me^{-m\beta\omega_0 }}$. By applying Eq.~(\ref{De1}), we further get
\begin{align}\label{DdD1}
	D^\dag(\zeta)\dif D(\zeta)=\frac{J_+\dif \zeta-J_-\dif\bar{\zeta}}{1+|\zeta|^2}+J_z\frac{\zeta\dif\bar{\zeta}-\bar{\zeta}\dif \zeta}{1+|\zeta|^2}.
\end{align}
Accordingly, the Uhlmann connection becoms
\begin{align}\label{AusHo2}
	&A_U=-\sum_{n\neq m}\frac{\chi_{nm}}{1+|\zeta|^2} \times \notag\\
	\quad &D(\zeta)
	|jn\rangle\langle jn|\left(J_+\dif \zeta-J_-\dif\bar{\zeta}\right)|jm\rangle\langle jm|D^\dag(\zeta)\notag\\
&=-\frac{\chi}{(1+|\zeta|^2)^2}\left[J_+(\dif\zeta+\zeta^2\dif\bar{\zeta})-J_-(\dif\bar{\zeta}+\bar{\zeta}^2\dif \zeta)\right.\notag\\
	&\left.+2J_z(\zeta\dif\bar{\zeta}-\bar{\zeta}\dif\zeta)\right],
\end{align}
where $\chi\equiv\chi_{n+1,n}=\chi_{n-1,n}=1-\text{\text{sech}}\frac{\beta\omega_0}{2}$. The $n=m$ terms have been included to simplify the expression since $\chi_{nn}=0$ gives no contributions.

Next, noting that $\dif \zeta=-\mi\dif \phi\zeta+\frac{1}{2}\dif \theta\me^{-\mi\phi}\sec^2\frac{\theta}{2}$, we obtain the final expression of $A_U$:
\begin{align}\label{AusHo5}
	A_U
	&=\mi\chi\left[\left(J_x\cos\phi+J_y\sin\phi\right)\cos\theta-J_z\sin\theta\right]\sin\theta\dif\phi\notag\\
	&\quad +\mi\chi\left(J_x\sin\phi-J_y\cos\phi\right)\dif\theta.
\end{align}
Using Eq.~(\ref{AusHo5}), the Uhlmann phase generated in a cyclic process can be evaluated.
The $A_U$ of the CSS only differs by a minus sign from that of the spin-$j$ system discussed in Ref.~\cite{OurPRA21}. Therefore, their Uhlmann phases only differ by a minus sign. As explained in Sec.~\ref{Sec.2}, this is because the CSS is related to the spin-$j$ system with a particular Hamiltonian.
Since the cases with $j=\frac{1}{2}$ and $1$ of the corresponding spin-$j$ systems have already been studied in Ref.~\cite{OurPRA21}, we will focus here on the $j=\frac{3}{2}$ CSS.
In Appendix \ref{app1}, we show that (I) the Uhlmann phase reduces to the Berry phase associated with the ground state as $T\rightarrow 0$, thereby confirming the correspondence explained in Ref.~\cite{P2}, and (II) the $T\rightarrow 0$ expressions of the CSSs agree with those of Ref.~\cite{Chaturvedi87} for pure states.

To explicitly show the properties of the Uhlmann phase of a
$j=\frac{3}{2}$ CSS, we choose the case evolving along the equator ($\theta=\frac{\pi}{2}$) in the parameter space.
In this case, the Uhlmann connection reduces to $A_U=-\mi\chi J_z \dif\phi$. The spin operators are \begin{align}\label{angularm}
J_x&=\left(
\begin{array}{cccc}
	0 & \frac{\sqrt{3}}{2} & 0 & 0 \\
	\frac{\sqrt{3}}{2} & 0 & 1 & 0 \\
	0 & 1 & 0 & \frac{\sqrt{3}}{2} \\
	0 & 0 & \frac{\sqrt{3}}{2} & 0 \\
\end{array}
\right),\notag\\
J_y&=\left(
\begin{array}{cccc}
	0 & \frac{\mi \sqrt{3}}{2} & 0 & 0 \\
	-\frac{\mi \sqrt{3}}{2} & 0 & \mi & 0 \\
	0 & -\mi & 0 & \frac{\mi \sqrt{3}}{2} \\
	0 & 0 & -\frac{\mi \sqrt{3}}{2}  & 0 \\
\end{array}
\right),\notag\\
J_z&=\left(
\begin{array}{cccc}
	-\frac{3}{2} & 0 & 0 & 0 \\
	0 & -\frac{1}{2} & 0 & 0 \\
	0 & 0 & \frac{1}{2} & 0 \\
	0 & 0 & 0 & \frac{3}{2} \\
\end{array}
\right).
\end{align}
We also assume that the evolution starts from $\phi=0$, so the initial density matrix is given by $\rho(0)=\frac{1}{Z} \mathrm{e}^{\frac{\mi\pi}{2}J_y} \me^{-\beta \omega_0 J_z} \mathrm{e}^{-\frac{\mi\pi}{2}J_y}$ according to Eq.~(\ref{rho1}). A straightforward evaluation shows that the Uhlmann phase generated in this process is
\begin{align}
	\theta_U &=\arg \operatorname{Tr} \left[ \frac{1}{Z}\me^{\beta \omega_0 J_x} \mathcal{P} \me^{\mi\chi \mathlarger{\oint}_{\gamma}  J_z \dif\phi} \right],
\end{align}
where we have used the fact $ \mathrm{e}^{\mi \theta J_y} J_z \mathrm{e}^{-\mi \theta J_y}	=-J_x \sin \theta+J_z \cos \theta$.

\begin{figure}[t]
	\centering
	\includegraphics[width=\columnwidth,clip]{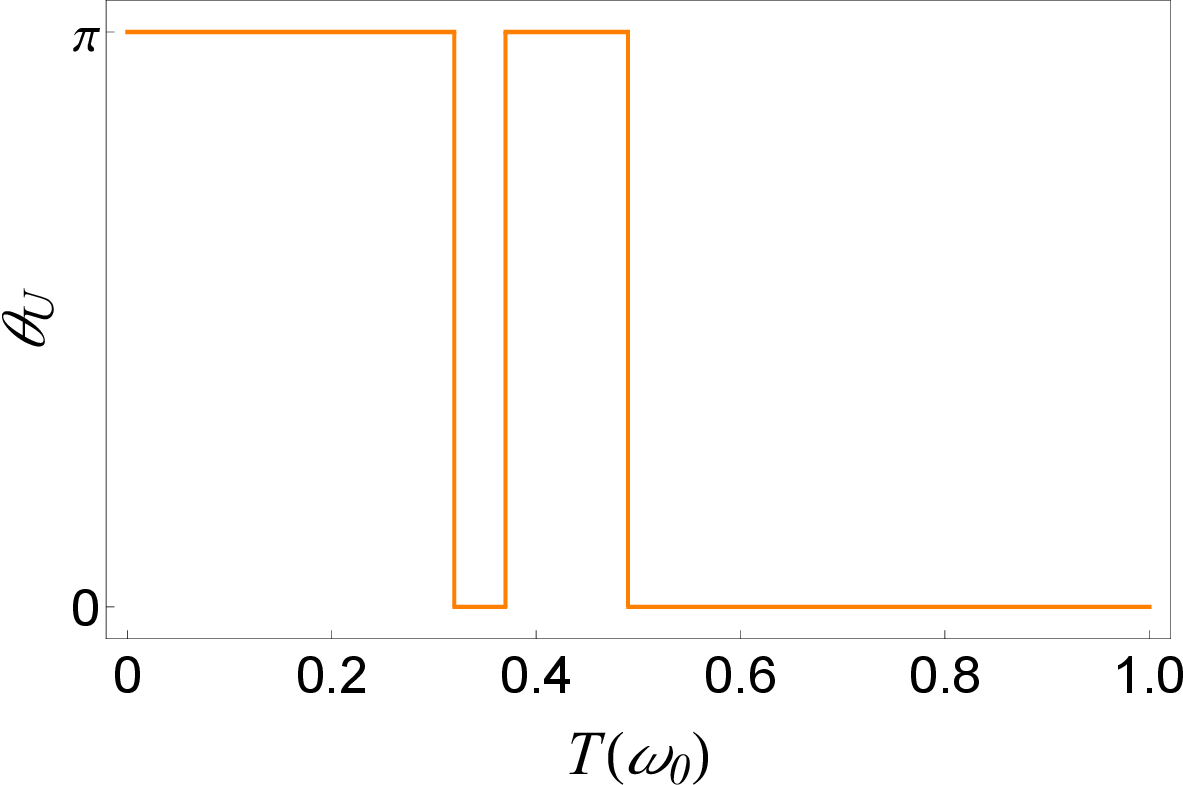}
	\caption{Uhlmann phase $\theta_{U}$ versus temperature for the CSS with $j=\frac{3}{2}$ along $\theta=\frac{\pi}{2}$.}
	\label{Fig1}
\end{figure}

We plot $\theta_U$ as a function of $T$ in Figure \ref{Fig1}. In the low temperature limit, $\theta_U=\pi$, which indicates that the topological properties of the system change after a cyclic evolution along the equator. More specifically, the final purification (in terms of the fiber in the bundle language) is ``antiparallel'' to the initial purification. As temperature increases and crosses the first critical temperature $T^1_c\approx 0.321\omega_0$, the value of $\theta_U$ suddenly drops to 0, implying that a topological change occurs in the evolution along the equator on the parameter space. The final purification now becomes ``parallel'' to the initial purification. As temperature continues to rise, subsequent topological transitions occur at $T^2_c\approx 0.376\omega_0$ and $T^3_c\approx 0.493\omega_0$. In the high temperature limit, $\theta_U$ remains zero, meaning the associated Uhlmann holonomy is topologically trivial. This is reasonable since $\lim_{T\rightarrow 0}\rho=\frac{1}{4}1_{4\times4}$. Therefore, the evolution loop $\rho(t)$ essentially becomes a single point with its horizontal lift $W(t)$  also reduced to a single point \cite{OurPRA21}. Hence, there is no difference between the initial and final fibers anymore. We mention that bosonic and fermionic coherent states at finite temperatures, in contrast, have been shown to have smooth behavior of the Uhlmann phase without any jumps~\cite{P2}.

\subsection{The interferometric geometric phase}
Here we first evaluate the IGP for the CSS following unitary evolution that satisfies the parallel-transport condition (\ref{pxcm1}). A natural candidate is the transformation that generates the CSS, as shown in Eq.~(\ref{rho1}). This yields
\begin{align}\label{IGPU}\rho(t)\equiv\rho(\zeta(t))=D(\zeta(t))\rho(0)D^\dag(\zeta(t)), \end{align}
where $\rho(0)=\frac{1}{Z}\me^{-\beta\hat{H}}$.
Initially, we set $\zeta(0)=0$, implying $\theta(0)=0$.
Using Eq.~(\ref{DdD1}), the parallel-transport condition (\ref{pxcm1}) can be satisfied by requiring
\begin{align}\label{ptca}
	&\operatorname{Tr}\left[\rho(0) D^{\dagger}(\zeta)\dif D(\zeta) \right]\notag\\
%	&= \frac{1}{Z}\operatorname{Tr}[\me^{-\beta\hat{H}}D^{\dagger}(\zeta)dD(\zeta)]\notag\\
%	&=\frac{1}{Z}\operatorname{Tr}[\me^{-\beta\hat{H}}(\frac{J_+\dif \zeta-J_-\dif\bar{\zeta}}{1+|\zeta|^2}+J_z\frac{\zeta\dif\bar{\zeta}-\bar{\zeta}\dif \zeta}{1+|\zeta|^2})]\notag\\
	=&\sum_{m} \frac{\me^{-m\beta\omega_0 }}{Z} \left\langle jm \right| \frac{J_+\dif \zeta-J_-\dif\bar{\zeta}+J_z(\zeta\dif\bar{\zeta}-\bar{\zeta}\dif \zeta)}{1+|\zeta|^2}\left| jm \right\rangle\notag\\
	%&=\frac{1}{Z(1+|\zeta|^2)} \sum_{m=-j}^{j} m\me^{-m\beta\omega_0 } (\zeta\dif\bar{\zeta}-\bar{\zeta}\dif \zeta) \notag\\
	=&\frac{2\mi}{Z(1+|\zeta|^2)} \sum_{m=-j}^{j} m\me^{-m\beta\omega_0 } \left(\tan^2\frac{\theta}{2}\dif\phi\right) \notag\\
	=&0.
\end{align}
A possible choice is $\tan^2\frac{\theta}{2}\dif\phi=0$, which implies either $\theta=0$ or $\dif \phi=0$. The former leads to $\zeta=0$, indicating that $D(\zeta)=1$, which is a trivial evolution. Therefore, we choose the latter and set $\phi$ to be constant, which corresponds to a longitude on the unit-sphere parameter space. We remark that the strengthened parallel-transport condition~\eqref{pxcm2} is also satisfied along the path.

\begin{figure}[t]
	\centering
	\includegraphics[width=3.0in,clip]{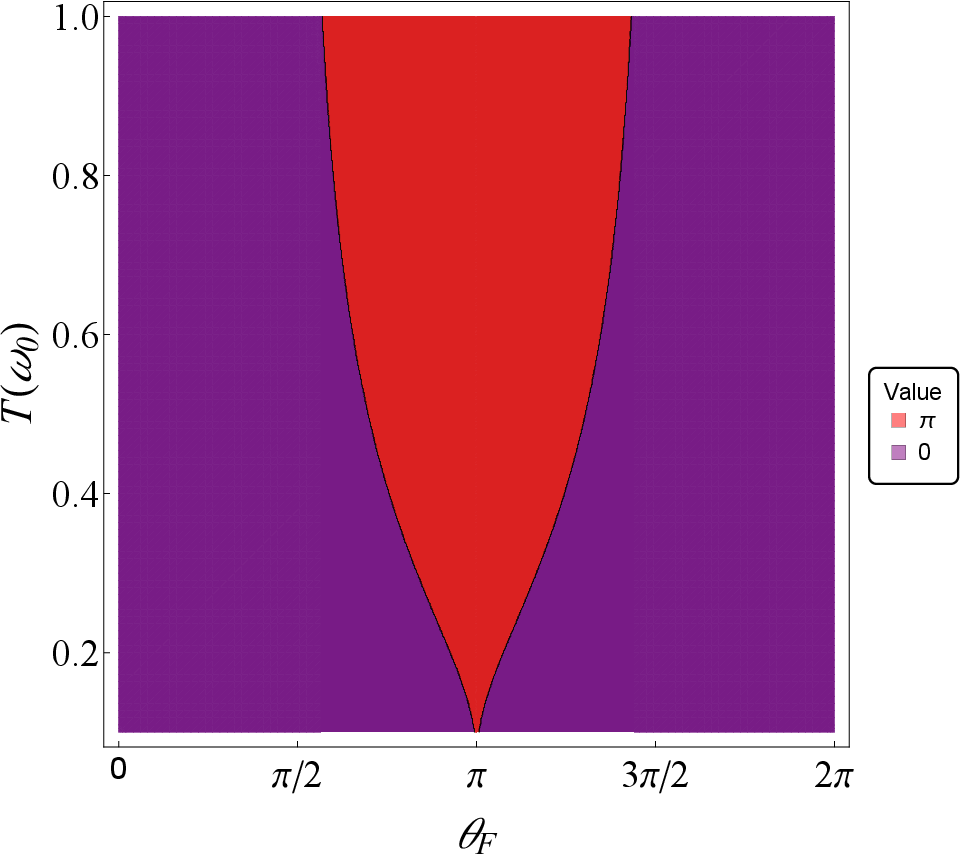}
	\includegraphics[width=2.9in,clip]{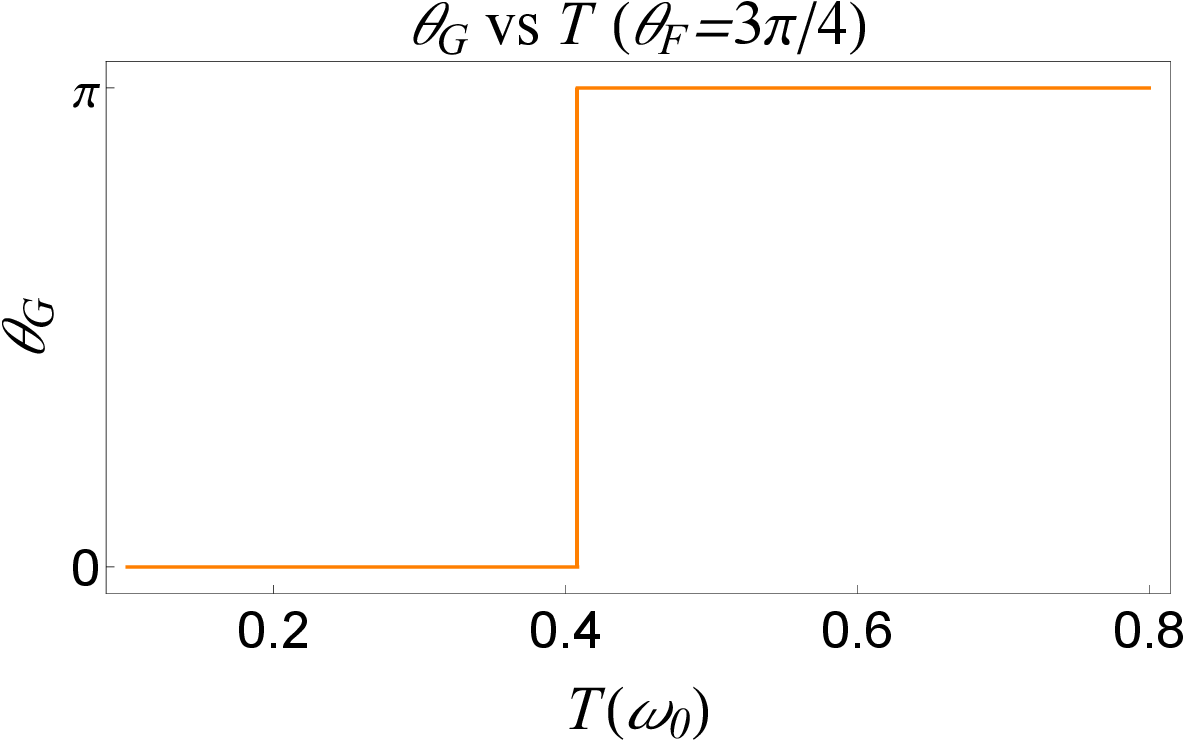}
	\caption{ (Top panel) Contour plot of the IGP $\theta_{G}$ versus temperature $T$ and $\theta_f$ for $j=\frac{3}{2}$. The geometric phase transitions when $\theta_f \in (\frac{\pi}{2}, \frac{3\pi}{2})$ are indicated by the boundaries separating the two values of $\theta_{G}$.
(Bottom panel) $\theta_{G}$ versus $T$ for the case with $\theta_f=\frac{3\pi}{4}$.
}
	\label{Fig2}
\end{figure}

To explicitly show the properties of the IGP along a path of fixed $\phi$, we set $j=\frac{3}{2}$ as the case of the Uhlmann phase and found
\begin{align}
	\theta_\text{G}(t)&=\arg \operatorname{Tr}[ \rho(0)D(t) ] \notag\\
	&=\arg \operatorname{Tr}[ \frac{1}{Z} \me^{-\beta\omega_0 J_z} \mathrm{e}^{\zeta(t) J_{+}} \mathrm{e}^{\ln \left(1+|\zeta(t)|^{2}\right) J_{z}} \mathrm{e}^{-\bar{\zeta}(t) J_{-}} ] \notag\\
	&=\arg \left[-2 e^{\beta \omega_0} \tan ^2\left(\frac{\theta(t) }{2}\right)+e^{2 \beta \omega_0}+1\right].
\end{align}
Interestingly, the result is independent of $\phi$. To visualize the behavior of the IGP, we introduce $\theta_f\equiv\theta(t_\text{final})$ as the final value of $\theta$ at the end of evolution by $D(t)$. In the top panel of Figure \ref{Fig2}, we present the contour plot of the IGP as a function of $T$ and $\theta_f$ for $j=\frac{3}{2}$. One observes that when $\theta_f \in (\frac{\pi}{2}, \frac{3\pi}{2})$, the value of the IGP undergoes a sudden jump from 0 to $\pi$ as the system crosses a critical temperature $T_c$. This means that $\rho(t_\text{final},T_c^-)$ is in phase with $\rho(0,T_c^-)$, whereas $\rho(t_\text{final},T_c^+)$ is out of phase with $\rho(0,T_c^+)$.

In the bottom panel of Figure \ref{Fig2}, we choose a special case with $\theta_f=\frac{3}{4}\pi$ to show the discontinuity of the IGP at $T_c=0.408\omega_0$. Similar to the case of the Uhlmann phase, this can be recognized as a geometric phase transition induced by temperature. The only difference is that, in general, the IGP is not directly related to the topological properties of the system. Furthermore, finite-temperature geometric phase transitions of the IGP seem to be less common than those of the Uhlmann phase. For a two-level Hermitian quantum system, a finite change in temperature in general does not induce any jump of the IGP \cite{Andersson_2016,Hou2023}. For a three-level system, a careful design of the evolution process can make it happen \cite{Hou2023}. Our results of the CSS thus illustrate additional transitions of the IGP and show that the CSS is suitable for studying interesting properties of mixed-state geometric phases.

\section{Mixed-state GEOMETRIC PHASES OF SSS}\label{Sec.4}
\subsection{One-Axis Squeezing}
\subsubsection{Uhlmann Phase}
For the SSS shown in Eq.~\eqref{Eq:SSS}, the parameter space is one-dimensional. Therefore, we need to determine the value of $\Theta$ to define a cyclic process for density matrices, which is essential for generating the Uhlmann phase. The density matrix of a system in thermal equilibrium governed by $H(\Theta)$ show in Eq.~\eqref{Eq:HTheta} is given by $\rho(\Theta)=\frac{1}{Z}\me^{-\beta H(\Theta)}$ with $Z=\operatorname{Tr}(\me^{-\beta H(\Theta)})$.
By examining Eq.~(\ref{SSS1}), we observe that the quadratic term of $J_x$ appears in the exponent, making it difficult to derive a general expression for the SSS. Therefore, for a concise illustration of the geometric phases of the SSS, we will focus our studies on a fixed value of $j$.
Since the behavior of the IGP for a two-level system ($j=\frac{1}{2}$) is relatively simple \cite{Andersson_2016}, we will begin with $j=1$ for our analysis. 
Crucially, Eq.~\eqref{AUE} indicates that the Uhlmann connection $A_U$ only depends on $\rho(t)$ and its eigenvalues and eigenstates.
Given that $\rho(t)=S(\Theta(t))\rho(0)S^\dag(\Theta(t))=\tilde{S}(\Theta(t))\rho(0) \tilde{S}^\dag(\Theta(t))$, the expressions following $S(\Theta)$ must be identical to those following $\tilde{S}(\Theta)$.
Consequently, the Uhlmann phase calculated above is the same as that from the evolution governed by $\tilde{S}(\Theta)$. The identification allows for a fair comparison with the IGP discussed later. We mention that the equivalent evaluation via $S(\Theta)$ simplifies the calculation without the distraction from the additional pure-state geometric phase associated with each $|jm\rangle$, which does not contribute to the final expression of the Uhlmann phase.

For the particular case of the $j=1$ SSS, the corresponding angular momentum operators are
\begin{align}\label{angumomen}
J_x&=\left(
	\begin{array}{ccc}
		0 & \frac{1}{\sqrt{2}} & 0 \\
		\frac{1}{\sqrt{2}} & 0 & \frac{1}{\sqrt{2}} \\
		0 & \frac{1}{\sqrt{2}} & 0 \\
	\end{array}
	\right),
J_y=\left(
\begin{array}{ccc}
	0 & \frac{\mi}{\sqrt{2}} & 0 \\
	-\frac{\mi}{\sqrt{2}} & 0 & \frac{\mi}{\sqrt{2}} \\
	0 & -\frac{\mi}{\sqrt{2}} & 0 \\
\end{array}
\right),\notag\\
J_z&=\left(
\begin{array}{ccc}
	-1 & 0 & 0 \\
	0 & 0 & 0 \\
	0 & 0 & 1 \\
\end{array}
\right).
\end{align}
From Eq.~\eqref{Eq:S}, we have
\begin{align}\label{Stheta}
S(\Theta)=\left(
\begin{array}{ccc}
	\frac{1}{2}+\frac{1}{2} \me^{-\frac{\mi \Theta}{2} } & 0 & -\frac{1}{2}+\frac{1}{2} \me^{-\frac{\mi \Theta}{2} } \\
	0 & \me^{-\frac{\mi \Theta}{2} } & 0 \\
	-\frac{1}{2}+\frac{1}{2}\me^{-\frac{\mi \Theta}{2}} & 0 & \frac{1}{2}+\frac{1}{2} \me^{-\frac{\mi \Theta}{2} } \\
\end{array}
\right).
\end{align}
A straightforward calculation shows that when $\Theta=4\pi$, $S(4\pi)=\mathbf{1}_{3\times3}$, which leads to $\rho(4\pi)=\rho(0)$ and gives a cyclic process.
The $4\pi$ periodicity also contrasts the $j=1$ SSS from an ordinary spin-1 state. Hereafter, we consider $\Theta\in[0,4\pi]$ for the $j=1$ SSS.

According to Eq.~(\ref{AUE}), the Uhlmann connection of the $j=1$ SSS is
\begin{align}
	A_U& = -\sum_{n \neq m} \chi_{nm} S \left|jn\right \rangle\left\langle j n \right| S^{\dagger} \dif S \left| j m \right\rangle \left\langle j m \right|S^{\dagger}\notag\\
&= \sum_{n \neq m} \chi S \left|j n\right \rangle\left\langle j n \right| \frac{\mi \left(J^2_+ +J^2_-\right)}{8}  \left| j  m \right\rangle \left\langle j  m \right|S^{\dagger} \dif \Theta \notag\\
&=\frac{\mi \chi}{4} \left[J^2_x-S J^2_y S^{\dagger} \right]\dif \Theta ,
\end{align}
where $\chi_{nm}=\frac{\mathrm{e}^{-\beta m  \omega_0}+\mathrm{e}^{-\beta n   \omega_0}-2 \mathrm{e}^{-\frac{\beta(m+n)  \omega_0}{2}}}{\mathrm{e}^{-\beta m \omega_0}+\mathrm{e}^{-\beta n  \omega_0}}$, and $\chi\equiv\chi_{n,n+2}=\chi_{n+2,n}=\frac{(\me^{\frac{\beta   \omega_0}{2}}-\me^{\frac{-\beta   \omega_0}{2}} )^2}{\mathrm{e}^{\beta   \omega_0}+\mathrm{e}^{-\beta   \omega_0}}$. In the last line, we have included the vanishing $n=m$ and $n=m-1$ terms. It is in general challenging to find a nontrivial evolution path along which $A_U$ is proportional to a constant matrix. The difficulty thus hinders the discovery of an analytical expression of the Uhlmann phase.

\begin{figure}[t]
	\centering
	\includegraphics[width=\columnwidth,clip]{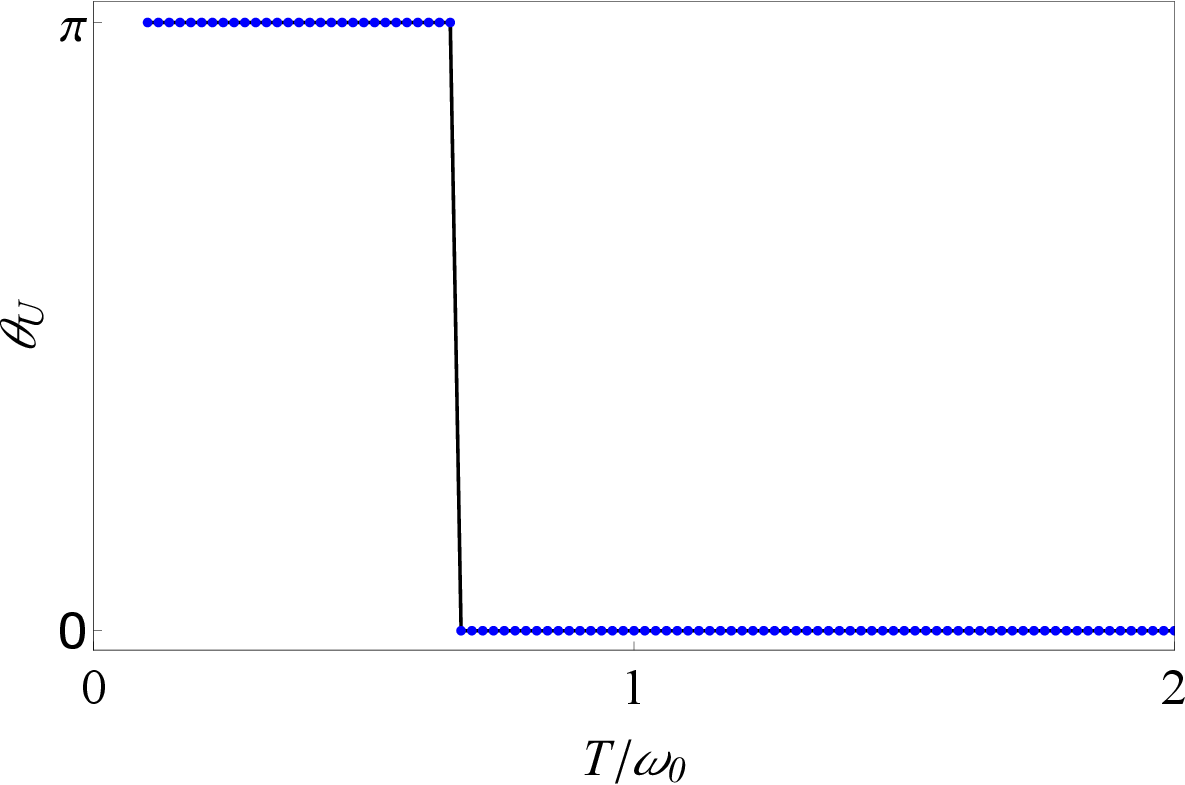}
	\caption{Uhlmann phase $\theta_{U}$ as a function of temperature $T$ for the $j=1$ one-axis SSS. A $\pi$-valued jump is observed at $T_c \approx 0.68 \omega_0$.}
	%The smooth curve indicates the absence of a transition in this case.T_c=  \me^{-0.38} \omega_0
	\label{Fig3}
\end{figure}

Nevertheless, numerical results can be obtained by using the Trotter-Suzuki approximation  \cite{Trotter59,Suzuki76} to tackle the path-ordered expression. Our findings are presented in Figure \ref{Fig3}, where $\theta_{U}$ is plotted as a function of $T$. The Uhlmann phase of SSS, similar to that of CSS, exhibits a $\pi$-valued jump at a finite temperature $T_c \approx 0.68 \omega_0$, indicating the existence of a temperature-induced topological phase transition. This result reveals that at finite temperatures, the initial and final purifications no longer remain parallel, and the ``angle'' between them undergoes an abrupt transition from ``parallel'' to ``antiparallel''. As shown in Fig.~\ref{Fig3}, this explains the jump in the Uhlmann phase from $\pi$ to $0$ at $T_c$. In the infinite-temperature limit, $\theta_U = 0$ again. This is reasonable since $\lim_{T\rightarrow +\infty}\rho$ is proportional to the identity matrix, and the associated horizontal lift shrinks to a single point, resulting in $\lim_{T\rightarrow +\infty}\theta_U = 0 \mod 2\pi$ \cite{P3}.

\subsubsection{The interferometric geometric phase}
As shown in the discussion around Eq.~(\ref{tildeSptc}), the operator $\tilde{S}$ automatically satisfies the strengthened parallel-transport condition of the IGP given by Eq.~(\ref{pxcm2}).
Accordingly, the IGP accumulated during this evolution is given by
\begin{align}\label{tildeIGP}
	\theta_\text{G}(t)=\arg \operatorname{Tr}[ \rho(0) \tilde{S}(t) ] =\arg  \left(\sum_{m}  \lambda_m \nu_m \me^{\mi \phi_m}\right),
\end{align}
where $\lambda_m$ is the $m$-th eigenvalue of the density matrix, $\nu_m=\langle m(0)| m(t) \rangle$ is the overlap between the initial and final eigenstates, and 
\begin{align}
 \phi_m &=\mi \int^{t}_{0} \dif t' \langle jm,\Theta(t') |\frac{\dif}{\dif t'} |jm,\Theta(t') \rangle \notag\\
  &=\mi \int^{t}_{0} \dif t' \langle jm| S^\dagger (t')\frac{\dif S(t')}{\dif t'} |jm\rangle.
 \end{align}
Using Eq.~\eqref{Stheta}, we finally obtain 
 \begin{align}\label{argtildeS}
	\theta_\text{G}(t)=\arg  \left[\frac{2 \cos \left(\frac{\Theta(t) }{4}\right) \cosh (\beta  \omega_0 )+1}{2 \cosh (\beta  \omega_0)+1}\right].
\end{align}

\begin{figure}[t]
	\centering	
    \includegraphics[width=\columnwidth,clip]{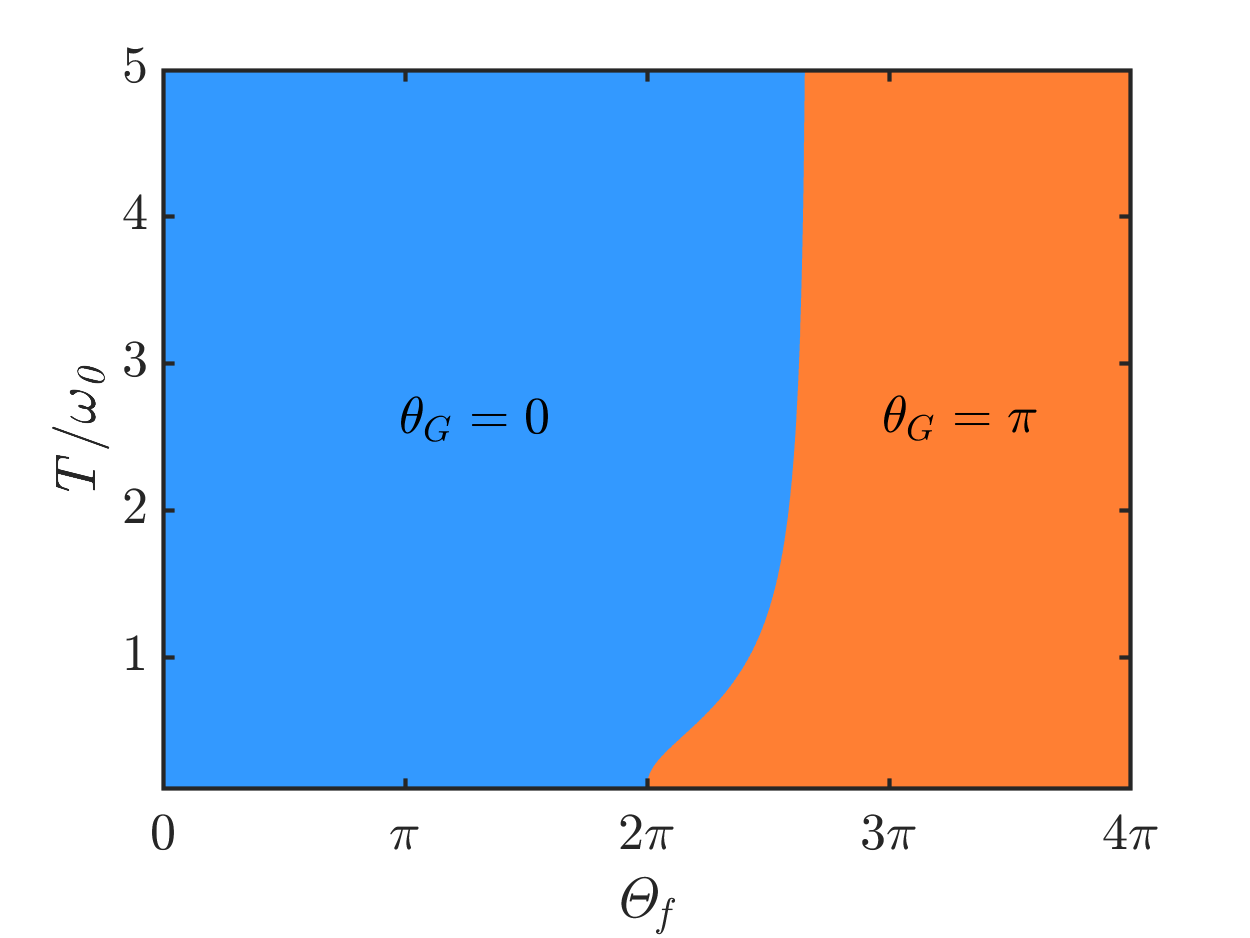}\\
	\caption{The IGP $\theta_\text{G}$ as a function of $T$ and $\Theta_f$ for the $j=1$ one-axis SSS, where $\Theta_f \in [0,4\pi]$. $\theta_\text{G}$ exhibits discontinuities with respect to both $\Theta_f$ and $T$. At finite temperatures, $\theta_\text{G}$ always experiences a $\pi$-jump when crossing $\Theta_f^c = 4 \arccos \left(- \operatorname{sech}  \left(\frac{\omega_0}{T}\right)/2\right)$, indicating a geometric phase transition. When $\Theta_f \in (2\pi,8\pi/3)$, there also exists a finite-temperature geometric phase transition as temperature varies.
	}
	\label{Fig11}
\end{figure}

To visualize the behavior of the IGP, we show in Fig.~\ref{Fig11}  $\theta_\text{G}$ as a function of temperature $T$ and the final squeezing parameter $\Theta_f$. Here $\Theta_f \equiv \Theta(t_{\text{final}})$ denotes the value of $\Theta$ at the end of the evolution governed by $\tilde{S}(t)$. As the system crosses a critical value $\Theta_f^c$ at finite temperatures, the IGP undergoes a sudden jump from 0 to $\pi$. In the zero-temperature limit, the critical squeezing parameter approaches $\Theta_f^c(T \to 0) = 2\pi$ while in the high-temperature limit it tends to $\Theta_f^c(T \to \infty) = \frac{8\pi}{3}$.
Similar to the previous example involving the CSS, the IGP according to the density matrix $\rho(t_\text{final}, T_c^-)$ remains in phase with $\rho(0, T_c^-)$, whereas that of $\rho(t_\text{final}, T_c^+)$ becomes out of phase with $\rho(0, T_c^+)$ for the $j=1$ one-axis SSS.
In general, the relation between $\Theta^c_f$ and $T_c$ for the $j=1$ one-axis SSS can be determined from the solutions of
\begin{align}
2 \cos \left(\frac{\Theta(t)}{4}\right) \cosh (\beta \omega_0) + 1 = 0,
\end{align}
as implied by Eq.~(\ref{argtildeS}). Explicitly, 
$\Theta^c_f = 4 \arccos \left(- \frac{1}{2} \operatorname{sech} \left(\frac{\omega_0}{T_c} \right) \right).$ 
Therefore, the IGP of the one-axis SSS can exhibit finite-temperature geometric phase transitions, analogous to those observed in the CSS. This highlights the rich and intriguing geometric structures inherent in the SSS.

\subsection{Two-Axis Squeezing}
Here we analyze the two-axis squeezing scenario and analyze its mixed-state geometric phases exemplified by the Uhlmann phase and IGP.

\subsubsection{Uhlmann Phase}
For the SSS given by Eq.~\eqref{Eq:tSSS} under two-axis squeezing, the parameter space becomes two-dimensional. The Uhlmann process can be generated by a cyclic path either on the complex plane or equivalently on the sphere parametrized by $(\theta, \phi)$ with $ z = \me^{-\mi \phi } \tan \left(\frac{\theta }{2}\right) $. The density matrix of a system in thermal equilibrium governed by the continuous transformation $K(z(t))$ of the two-axis SSS at temperature $T$ then defines a loop $ C(t) := \rho(z(t)) $ in the manifold of density matrices. Explicitly,
\begin{align}\label{rhotwoaxis}
	\rho(z)=\frac{1}{Z}\me^{-\beta\hat{H}(z)}=\frac{1}{Z}K(z)\me^{-\beta\hat{H}}K^\dag(z),
\end{align}
where $\hat{H}=\me^{-\beta\omega_0 \hat{J}_z}$.

We continue to focus on the case of the $ j = 1 $ SSS for comparison. The corresponding angular momentum operators are given in Eq.~\eqref{angumomen}. From Eq.~\eqref{Eq:K}, we obtain
\begin{align}
K(\theta,\phi)=\left(
\begin{array}{ccc}
	K_{11} & 0 & K_{13} \\
	0 & 1 & 0 \\
	-K_{13} & 0 & K_{11} \\
\end{array}
\right),
\end{align}
where \begin{align} K_{11}& = \cos\left[2 \tan \left(\frac{\theta}{2}\right)\right] ,\notag\\ K_{13}& = \frac{\mi \me^{\mi \phi}}{2} \left[ \me^{2 \mi \tan \left(\frac{\theta}{2}\right)} - \me^{-2 \mi \tan \left(\frac{\theta}{2}\right)} \right].\end{align}
A relatively simple path can be chosen along the equator, i.e., $ \theta = \frac{\pi}{2} $, where $ \phi $ varies with time. We assume the evolution starts from $ \phi = 0 $, so the initial density matrix is given by
$\rho(0) = \frac{1}{Z} \mathrm{e}^{J^2_+ - J^2_-} \me^{-\beta \omega_0 J_z} \mathrm{e}^{J^2_- - J^2_+}$
according to Eq.~(\ref{rhotwoaxis}).

The Uhlmann connection for the two-axis squeezing of the $ j=1 $ SSS can be obtained in similar way following that of the one-axis squeezing case. Based on Eq.~(\ref{AUE}),
\begin{align}\label{twoaxisAu}
A_U = -\sum_{n \neq m} \chi_{nm} K \left| jn \right\rangle \left\langle jn \right| K^\dagger \dif K \left| jm \right\rangle \left\langle jm \right| K^\dagger.
\end{align}
However, the main difference lies in the complexity of $ K(z) = \me^{z J^2_+ - \bar{z} J^2_-} $, which can be difficult to disentangle. As a result, obtaining an analytical expression for $ K^\dagger \dif K $ is nontrivial. Fortunately, since we are only interested in the case with $ j = 1 $, $ K^\dagger \dif K $ can be represented by a $ 3 \times 3 $ matrix in the $ |jn\rangle \langle jm| $ basis:
\begin{align}
K^\dagger \dif K = \mi \left(
\begin{array}{ccc}
	- \sin^2(2) & 0 & -\frac{ \me^{\mi \phi} \sin(4)}{2} \\
	0 & 0 & 0 \\
	-\frac{ \me^{-\mi \phi} \sin(4)}{2} & 0 & \sin^2(2) \\
\end{array}
\right) \dif \phi.
\end{align}
To compute $ A_U $, a crucial step is to determine the coefficients $ \chi_{nm} $. We find that all nonzero contributions to $ A_U $ are governed by
\begin{align}
\chi \equiv \chi_{n, n+2} = \frac{ \left( \me^{\frac{\beta \omega_0}{2}} - \me^{- \frac{\beta \omega_0}{2}} \right)^2 }{ \me^{\beta \omega_0} + \me^{- \beta \omega_0} },
\end{align}
which follows from eliminating the diagonal elements of $ K^\dagger \dif K $.
Eventually, we obtain the Uhlmann connection $ A_U $ corresponding to the specific evolution path $ C(t) $:
\begin{align}
	A_U= \mi \left(
	\begin{array}{ccc}
		-\frac{\sin ^2(4)}{2} & 0 & \frac{\sin (8) e^{ \mi \phi } }{4}  \\
		0 & 0 & 0 \\
		\frac{\sin (8) e^{-\mi \phi } }{4} & 0 & \frac{\sin ^2(4)}{2} \\
	\end{array}
	\right)\dif \phi.
\end{align}

\begin{figure}[t]
	\centering
	\includegraphics[width=\columnwidth,clip]{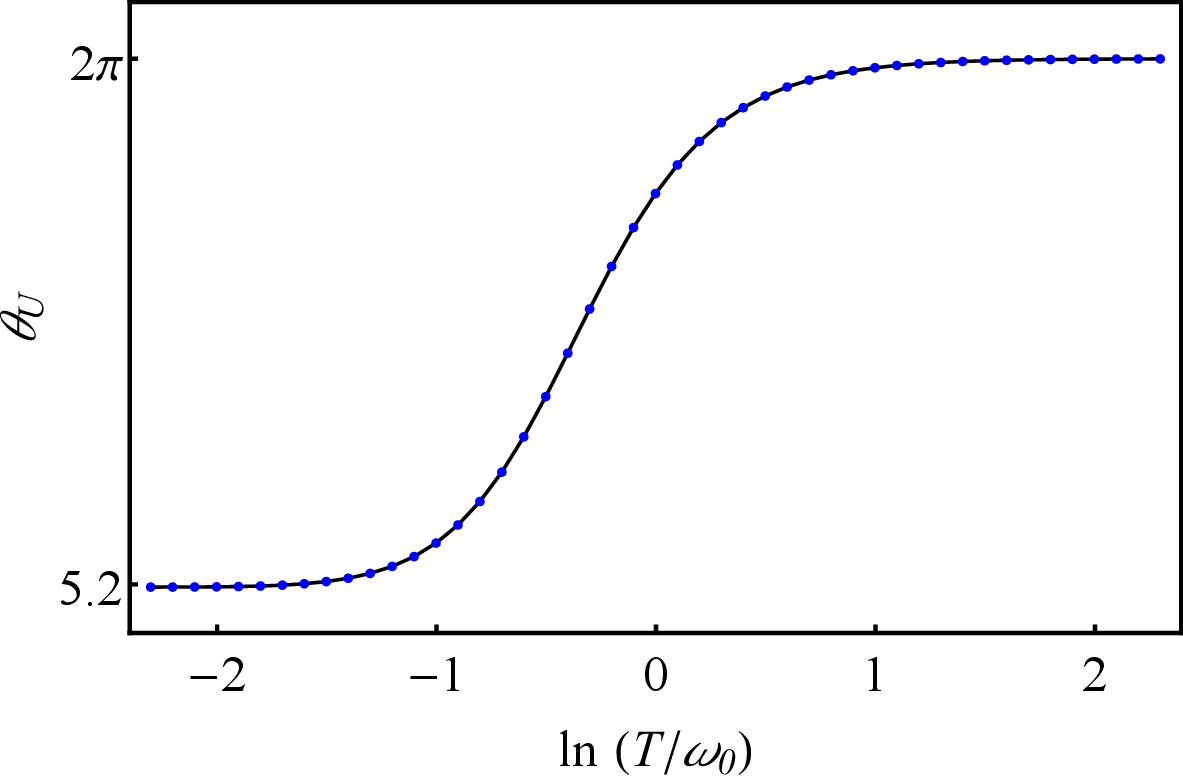}
	\caption{Uhlmann phase $\theta_{U}$ of the $j=1$ two-axis SSS as a function of temperature $T$ on semi-log scale. The smooth curve indicates the absence of a transition.}
	\label{Fig8}
\end{figure}

By applying the Trotter-Suzuki approximation once again, our result of the Uhlmann phase as a function of temperature $ T $ on a semi-logarithmic scale is shown in Figure~\ref{Fig8}. The smooth curve indicates that at finite temperatures, the ``angle" between the initial and final purifications no longer jump between $0$ and $\pi$ but varies continuously with temperature. This behavior is qualitatively different from the cases discussed earlier. Our results demonstrate that different quantum processes can exhibit different topological characteristics. The Uhlmann holonomy shown by the Uhlmann phase can be different even though the CSS, one-axis SSS, and two-axis SSS all follow unitary processes. In the infinite-temperature limit, we again find $ \theta_U = 2\pi \equiv 0 \mod 2\pi $ for the two-axis squeezing, which is consistent with the Uhlmann phases of the CSS and one-axis SSS.

\begin{figure}[t]
	\centering
	\includegraphics[width=3.2in,clip]{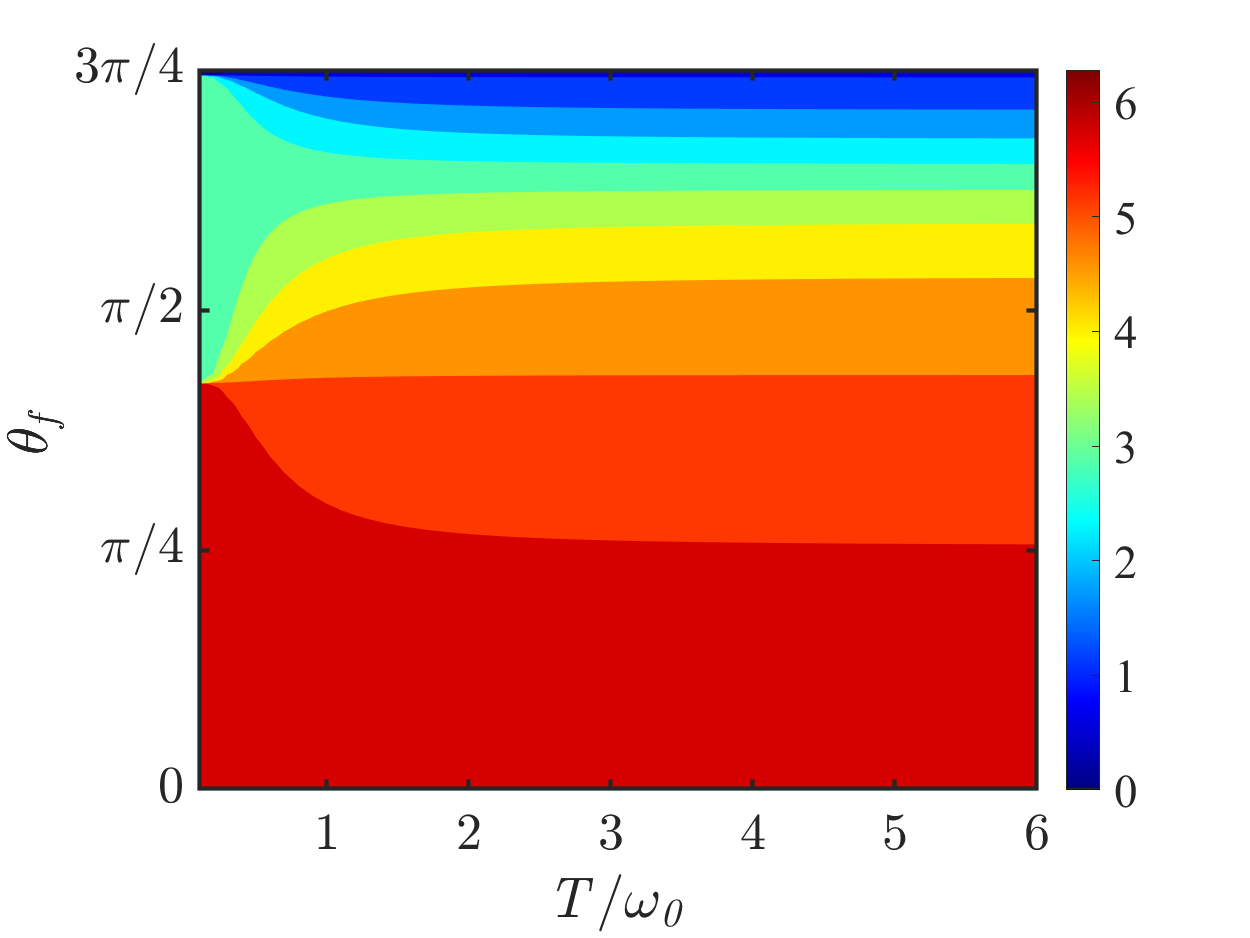}\\
	\includegraphics[width=1.68in,clip]{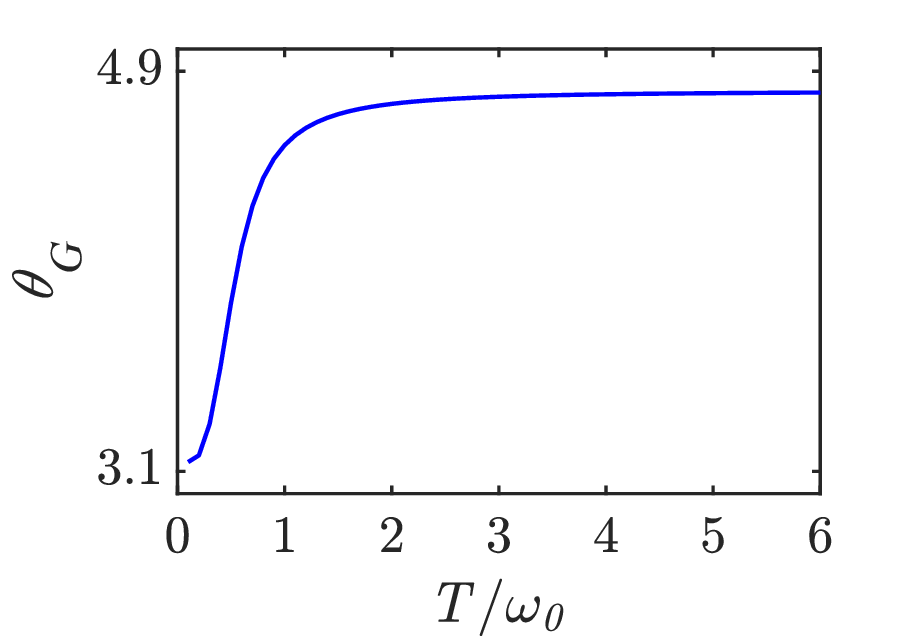}
	\includegraphics[width=1.68in,clip]{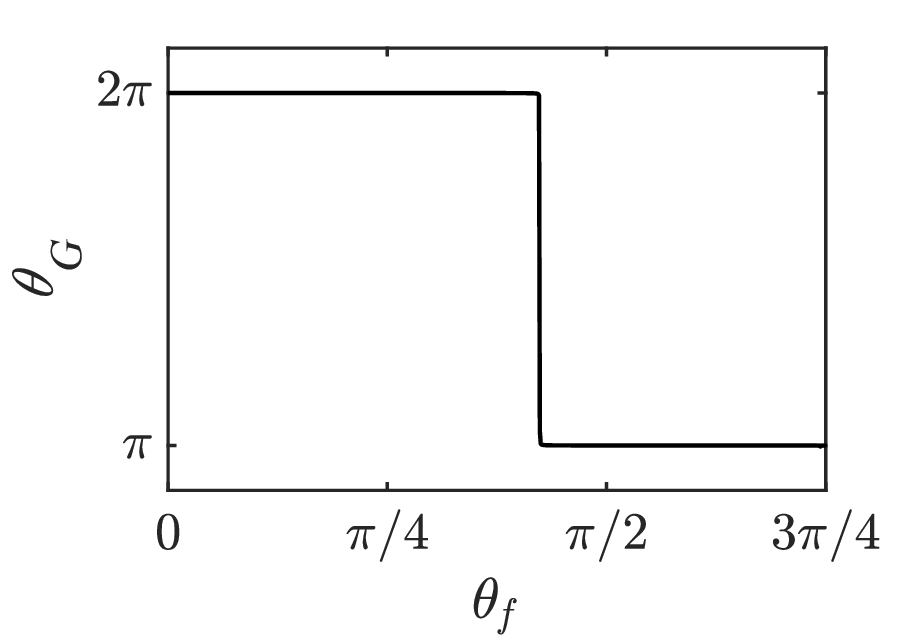}
	\caption{(Top panel) Contour plot of the IGP $\theta_\text{G}$ of the $j=1$ two-axis SSS as a function of $T$ and $\theta_f$, where $\theta_f \in [0,\frac{3\pi}{4}]$. $\theta_\text{G}$ is continuous with respect to both $\theta_f$ and $T$ (apart from a jump at $T=0$). (Bottom panels) $\theta_\text{G}$ vs $T$ at $\theta_f=\pi/2$, where $\theta_\text{G}$ exhibits no jumps (left), and $\theta_\text{G}$ vs $\theta_f$ at $T=0$, where there is a $\pi$-jump at $\theta_f \approx 1.33$rad (right). } 
	\label{Fig9}
\end{figure}

\subsubsection{The IGP}
For the two-axis SSS, we find that the strengthened parallel transport condition~\eqref{pxcm2} of the IGP holds for a nontrivial process which starts from the pole at $\theta = 0$ and proceeds along the meridian with $\phi = \frac{\pi}{2}$ under the transformation $K(\theta, \phi)$. Of course, such a path also respects the condition~\eqref{pxcm1}. The density matrix $\rho(t)$ of this unitary evolution is
\begin{align}\label{rhotwoaxis2}
	\rho(z(t))=K(z(t)) \rho(0) K^\dag(z(t)),
\end{align}
where $\rho(0)=\frac{\me^{-\beta\omega_0 J_z}}{Z}$.
In the $j=1$ representation, the matrix form of $K^\dagger \dif K$ is
\begin{align}
	K^\dagger \dif K=- \mi \left(
	\begin{array}{ccc}
		0 & 0 & \frac{2}{\cos (\theta )+1} \\
		0 & 0 & 0 \\
		\frac{2}{\cos (\theta )+1} & 0 & 0 \\
	\end{array}
	\right)\dif \theta ,
\end{align}
where all diagonal elements are zero. Therefore, the IGP parallel-transport condition \eqref{pxcm2} is equivalent to
\begin{align}\label{eqref19} \langle jm | K^\dagger(t) \dot{K}(t) | jm \rangle = 0, \quad m = -1, 0, 1, \end{align}
and is always satisfied. The weaker parallel transport condition~\eqref{pxcm1} is also satisfied. In this case, the IGP is given by
\begin{align}
	\theta_\text{G}(t)&=\arg \operatorname{Tr}[ \rho(0)K(t) ] \notag\\
	&=\arg \operatorname{Tr}[ \frac{1}{Z} \me^{-\beta\omega_0 J_z} \me^{z J^2_+ -\bar{z} J^2_-}] \notag\\
	&=\arg \left[\frac{2 \cos \left(2 \tan \left(\frac{\theta }{2}\right)\right)}{\text{sech}(\beta \omega_0)+2}+ \frac{1-2 \mi \sin \left(2 \tan \left(\frac{\theta }{2}\right)\right)}{2 \cosh (\beta \omega_0)+1}\right].
\end{align}

Since $\tan\left(\frac{\theta}{2}\right) \to \infty$ as $\theta \to \pi$ causes rapid oscillations in the values of $\theta_{\text{G}}(t)$, the range of $\theta$ is restricted to $[0, \frac{3}{4}\pi]$ in our numerical calculation to ensure accurate results. In the top panel of Figure \ref{Fig9}, the IGP $ \theta_\text{G} $ is plotted as a function of temperature $ T $ and the two-axis squeezing parameter $ \theta_f \equiv \theta(t_\text{final}) $. In this two-axis squeezing model, similar to the Uhlmann phase, there is no temperature-induced geometric phase transition. In the bottom panel (left side), we show the behavior of $ \theta_\text{G} $ as a function of $ T $ for $ \theta_f = \pi/2 $, where $ \theta_\text{G} $ continuously increases with $ T $. At finite temperatures, similar to the non-adiabatic IGP of the one-axis squeezing of the SSS, we find that $ \theta_\text{G} $ also changes continuously with $ \theta_f $ without any abrupt jumps.

However, in the zero-temperature limit, $ \theta_\text{G} = \arg \left[ \cos \left( 2 \tan \left( \frac{\theta}{2} \right) \right) \right] $, which suggests that a $ \pi $-jump in the IGP may occur at the zeros of $ \cos \left( 2 \tan \left( \frac{\theta}{2} \right) \right) $, corresponding to sign changes within the function’s argument. Therefore, as $ \theta_f $ crosses $ \theta_f = 2 \tan^{-1}\left( \frac{\pi}{4} \right) \approx 1.33 \, \text{rad} $, the value of $ \theta_\text{G} $ undergoes a $ \pi $-jump (from $ 2\pi $ to $ \pi $ or vice versa), signaling a geometric phase transition. This behavior of $ \theta_\text{G} $ is shown in the bottom right panel of Figure \ref{Fig9}.
For the $ j=1 $ two-axis SSS, however, no finite-temperature jumps occur for the mixed-state geometric phases studied here.

\section{Implications}\label{Sec.5}
Since the CSS and SSS have been realized in condensed matter or atomic, molecular, and optical  systems~\cite{Est_ve_2008,Gross_2010,Huang_2021,PhysRevA.109.022438,Hosten_2016}, experimental generation and measurement of the mixed-state geometric phases may become feasible in the future. However, imposing the parallel-transport condition for the IGP or Uhlmann phase requires precise controls of the systems and the ancilla representing the environment, which can be challenging for natural systems in laboratories.

On the other hand, the rapid development of quantum computers opens possibilities of simulating complex quantum systems. By using an ancilla qubit acting as the environment and coupling it to a system qubit, Ref.~\cite{npj18} shows that the Uhlmann phase of a spin-$1/2$ system can be simulated on a quantum computer by using two engineered Hamiltonians to respectively evolve the system and ancilla simultaneously. Since a spin-$j$ object may be decomposed as a collection of $2j$ spin-$1/2$ objects, one may use $2j$ qubits as a system of a spin-$j$ system coupled to an ancilla system of $2j$ qubits to realize the CSS or SSS coupled to the environment. It then follows a series of technical challenges to construct suitable engineered Hamiltonians for the system and ancilla in order to generate the corresponding IGP or Uhlmann phase when the corresponding parallel-transport condition is imposed during the evolution. After the targeted geometric phase is generated on a quantum computer, one may follow Ref.~\cite{npj18} to couple additional measurement qubits to the composite system and perform state tomography to extract the phase accumulated during the constrained evolution.

\section{Conclusion}\label{Sec.6}
We have derived the Uhlmann phase and IGP  for the CSS and SSS. Selected examples are fully solved to demonstrate interesting properties of those mixed-state geometric phases. For the $j=3/2$ CSS, the Uhlmann phase exhibits topological phase transitions characterized by abrupt, quantized jumps while the IGP shows geometric phase transitions with discontinuous jumps as temperature varies. For the $j=1$ one-axis SSS, both Uhlmann phase and IGP display similar temperature-dependent transitions. In contrast, the $j=1$ two-axis SSS shows different behavior as both Uhlmann phase and IGP vary smoothly with temperature, suggesting the absence of temperature-induced phase transitions. These findings highlight the rich interplay between temperature, spin properties, and geometric phases, and offer insights into topological and dynamic behavior of quantum systems at finite temperatures. Furthermore, our results pave the way for potential experimental explorations and numerical simulations to probe interesting spin states with implications for quantum metrology, quantum information processing, and the study of topological and geometric quantum phase transitions.

\section*{ACKNOWLEDGEMENTS}
H.G. was supported by the Innovation Program for Quantum Science and Technology (Grant No. 2021ZD0301904) and the National Natural Science Foundation of China (Grant No. 12074064). X.Y.H. was supported by  the National Natural Science Foundation of China (Grant No. 12405008) and the Jiangsu Funding Program for Excellent Postdoctoral Talent (Grant No. 2023ZB611). C.C.C. was supported by the NSF (No. PHY-2310656) and DOE (No. DE-SC0025809).

\appendix
\section{Correspondence between Uhlmann phase and Berry phase of CSS as $T\rightarrow 0$}\label{app1}
To evaluate the Uhlmann phase of the CSS, we introduce $g(t)=\mathcal{P}\me^{-\mathlarger{\int}_{0,\gamma}^t A_U(t')}$, which is an operator depending on the evolution curve $\gamma$. Since $\gamma(\tau)=\gamma(0)$, $g(\tau)$ is in fact the Uhlmann holonomy. The operator $g(t)$ satisfies the differential equation
\begin{align}\label{dg}
	\frac{\dif g(t)}{\dif t}=\frac{\chi}{1+|\zeta|^2}D(\zeta(t))(J_+\dot{\zeta}-J_-\dot{\bar{\zeta}})D^\dag(\zeta(t))g(t)
\end{align}
subject to the initial condition $g(0)=1$.
To solve this equation, we define $g'(t)=D^\dag(\zeta(t))g(t)$. By taking the complex conjugation of both sides of Eq.~(\ref{DdD1}), we get
\begin{align}\label{dDD1}
	\dot{D}^\dag D=\frac{1}{1+|\zeta|^2}\left[J_-\dot{\bar{\zeta}}-J_+\dot{\zeta}+J_z\left(\bar{\zeta}\dot{\zeta}-\zeta\dot{\bar{\zeta}}\right)\right].
\end{align}
Using Eqs.~(\ref{dg}) and (\ref{dDD1}), we have
\begin{align}
	&\frac{\dif g'(t)}{\dif t}=\dot{D}^\dag D g'(t)+\frac{\chi}{1+|\zeta|^2}\left(J_+\dot{\zeta}-J_-\dot{\bar{\zeta}}\right)g'(t)\notag\\
	&=\frac{1}{1+|\zeta|^2}\left[-\eta\left(J_+\dot{\zeta}-J_-\dot{\bar{\zeta}}\right)+J_z\left(\bar{\zeta}\dot{\zeta}-\zeta\dot{\bar{\zeta}}\right)\right]g'(t),\notag
\end{align}
where $\eta=\text{\text{sech}}\frac{\beta\omega_0}{2}$. Solving this equation, we obtain
\begin{align}\label{gtau}
	g(\tau)=&D(\zeta(\tau)) \mathcal{P}\me^{\frac{\mathlarger{\oint}_\gamma\left[-\eta\left(J_+\dif\zeta-J_-\dif\bar{\zeta}\right)+J_z\left(\bar{\zeta}\dif\zeta-\zeta\dif\bar{\zeta}\right)\right]}{1+|\zeta|^2}}D^\dag(\zeta(0)).
\end{align}
Therefore, the Uhlmann phase of the CSS is
\begin{align}\label{ThetaU}
	\theta_U&=\arg\text{Tr}\left[\rho(\zeta(0))g(\tau)\right]\notag\\
	&=\arg\text{Tr}\left[\rho(0)\mathcal{P}\me^{\frac{\mathlarger{\oint}_\gamma\left[-\eta\left(J_+\dif\zeta-J_-\dif\bar{\zeta}\right)+J_z\left(\bar{\zeta}\dif\zeta-\zeta\dif\bar{\zeta}\right)\right]}{1+|\zeta|^2}}\right],
\end{align}
where $\rho(0)=\frac{1}{Z}\me^{-\beta \omega_0 J_z}$. As $T\rightarrow 0$, $\lim_{\beta\rightarrow \infty}\eta=\lim_{\beta\rightarrow \infty}\text{\text{sech}}\frac{\beta\omega_0}{2}=0$, and $\rho(0)\approx |j,-j\rangle\langle j,-j|$, then
\begin{align}\label{ThetaUT0}
	\lim_{T\rightarrow 0}\theta_U&=\arg\langle j,-j|\mathcal{P}\me^{\frac{1}{1+|\zeta|^2}\mathlarger{\oint}_\gamma J_z\left(\bar{\zeta}\dif\zeta-\zeta\dif\bar{\zeta}\right)}|j,-j\rangle\notag\\
	&=\mi \frac{j}{1+|\zeta|^2}\oint_\gamma \left(\bar{\zeta}\dif\zeta-\zeta\dif\bar{\zeta}\right),
\end{align}
where $J_z|j,-j\rangle=-j|j,-j\rangle$ has been applied. This result precisely reduces to the Berry phase of the ground state $|j,-j\rangle$~\cite{Chaturvedi87}. More specifically, when $j=\frac{3}{2}$ and $\theta=\frac{\pi}{2}$, $|\zeta|=1$ and thus $\theta_U=\mi \frac{3}{4} \mathlarger{\oint}_\gamma (-2\mi \tan^2\frac{\theta}{2}) \dif\phi= 3\pi \equiv \pi \mod 2\pi$, which confirms the result shown in Fig.~\ref{Fig1} as $T\rightarrow 0$.

%\bibliographystyle{apsrev}
%\bibliography{Review1}

\end{document}